\documentclass{ws-procs9x6}
\usepackage{epsfig}
\newcommand{\gp}{$\vec \gamma p \rightarrow p\pi^+\pi^-${ }}

\begin{document}

\title{Helicity-Dependent Angular Distributions for Double-Charged-Pion
Photoproduction\footnote{\uppercase{T}his work was supported by the
\uppercase{U.S.}~\uppercase{D}epartment of \uppercase{E}nergy under
grant \uppercase{DE--FG02--95ER40901}. \uppercase{S}outheastern
\uppercase{U}niversities \uppercase{R}esearch \uppercase{A}ssociation
(\uppercase{SURA}) operates the \uppercase{T}homas
\uppercase{J}efferson \uppercase{N}ational \uppercase{A}ccelerator
\uppercase{F}acility under \uppercase{U.S.}~\uppercase{D}epartment of
\uppercase{E}nergy contract \uppercase{DE--AC05--84ER40150}.}}

\author{S.~Strauch for the CLAS Collaboration}

\address{Department of Physics\\
The George Washington University\\
Washington, D.C. 20052, USA\\ 
E-mail: strauch@gwu.edu}

\maketitle

\abstracts{ Two-pion photoproduction in the reaction \gp has been
studied at Jefferson Lab Hall B using a circularly-polarized tagged
photon beam in the energy range between 0.6~GeV and 2.3~GeV.
Beam-helicity-dependent angular distributions of the final-state
particles were measured. The large cross-section asymmetries that have
been found exhibit strong sensitivity to the kinematics of the
reaction, and are compared with preliminary model calculations by
Mokeev and Roberts.}

\section{Introduction}

Many nucleon resonances in the mass region above 1.6~GeV decay
predominantly through $\Delta\pi$ or $N\rho$ intermediate states into
$N\pi\pi$ final states (see the Particle-Data Group
review\cite{pdg02}). This makes electromagnetic exclusive double-pion
production an important tool in the investigation of $N^*$ structure
and reaction dynamics, as well as in the search for ``missing'' baryon
states. Unpolarized cross-section measurements of double-pion
electroproduction have been reported recently by the CLAS
collaboration.\cite{Ripani03} Further constraints are to be found in
polarization observables.

Here, for the first time in the resonance region, a measurement of the
\gp reaction is reported, where the photon beam is circularly
polarized and no nuclear polarizations (target or recoil) are
specified.  The cross-section asymmetry is defined by:
\begin{equation}
   A = \frac{1}{P_\gamma} \cdot 
   \frac{\sigma^+ - \sigma^-}{\sigma^+ + \sigma^-}\;,
\end{equation}
where $P_\gamma$ is the degree of circular polarization of the photon
and $\sigma^\pm$ is the cross section for the two photon-helicity
states $\lambda_\gamma=\pm 1$. For this kind of study, a final state
of at least three particles is necessary, since reactions with
two-body final states are always coplanar and have identical cross
sections for unpolarized or circularly polarized photons, so that $A =
0$.

\section{Experiment}

\begin{figure}[h!]
\centerline{\epsfig{file=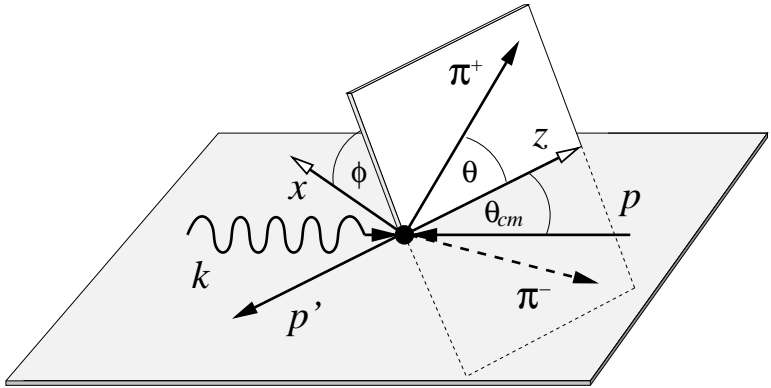,width=4.0in}}
\caption{Angle definitions for the circular polarized real-photon
reaction \gp in the helicity frame; $\theta_{cm}$ is defined in the
overall center-of-mass frame, $\theta$ and $\phi$ are defined as the
$\pi^+$ polar and azimuthal angles in the rest frame of the
$\pi^+\pi^-$ system.}
\label{fig:def}
\end{figure}

The \gp reaction was studied with the CEBAF Large Acceptance
Spectrometer (CLAS)\cite{Mecking03} at Jefferson Lab.  A schematic
view of the reaction is shown in Fig.~\ref{fig:def}.%
\footnote{The definition of $\phi$ is following the convention of
Schilling, Seyboth and Wolf\cite{Schilling70}, and differs by a phase
of $\pi$ from $\Phi^*$ in Ref.~\cite{Strauch03}.}
Longitudinally polarized electrons with an energy of 2.4~GeV were
incident on the thin radiator of the Hall-B Photon
Tagger\cite{Sober00} and produced circularly-polarized tagged photons
in the energy range between 0.6~GeV and 2.3~GeV. The collimated photon
beam irradiated a liquid-hydrogen target. The circular polarization of
the photon beam was determined from the electron-beam polarization and
the ratio of photon and incident electron energy.\cite{Maximon59} The
reaction channel was identified by the missing-mass technique, which
requires the detection of at least two out of three final-state
particles ($p$, $\pi^+$, and $\pi^-$).  Owing to the large angular
acceptance of the CLAS, complete azimuthal angular distributions of
the cross-section asymmetries were observed.

\section{Results}

\begin{figure}[h!]
\centerline{\epsfig{file=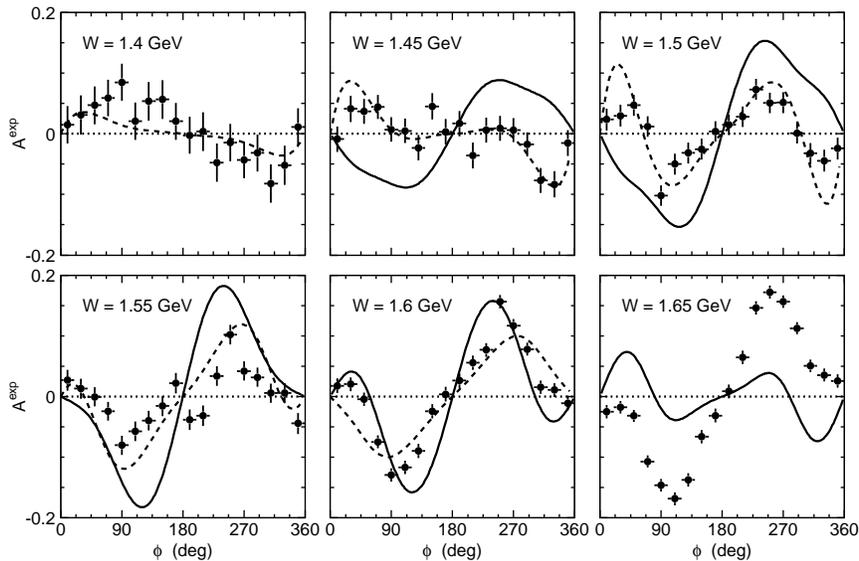,width=\textwidth}}
\caption{Preliminary angular distributions for six different
center-of-mass energy bins ($\Delta W = 25$ MeV) of the cross-section
asymmetry for the \gp reaction. The dashed curves are calculations by
Roberts\protect\cite{Roberts97} ($4\pi$ integrated, $W \le
1.60$~GeV). The solid curves are calculations by Mokeev {\it et
al.}\protect\cite{Mokeev01} (acceptance corrected, $W \ge 1.45$~GeV).}
\label{fig:mokeev_roberts}
\end{figure}

The \gp reaction has been analyzed for center-of-mass energies $W$ up
to 2.3~GeV. Figure \ref{fig:mokeev_roberts} shows preliminary $\phi$
angular distributions of the cross-section helicity asymmetry for
various selected 25-MeV wide c.m. energy bins between $W = 1.40$~GeV
and $1.65$~GeV. The data are integrated over the full CLAS
acceptance. The preliminary analysis shows large asymmetries, with the
symmetry $A(\phi)=-A(2\pi-\phi)$. This is expected from parity
conservation.\cite{Schilling70} A more detailed analysis has revealed
a rich structure of these data with rapid changes of the angular
distributions with photon energy or with any other kinematical
variable.\cite{Strauch03}

K.~Schilling, P.~Seyboth and G.~Wolf discussed the case of
photoproduction of vector mesons by polarized photons on an
unpolarized nucleon and their subsequent decay
distribution.\cite{Schilling70} Preliminary calculations of
cross-section asymmetries in the general case of \gp were done by Oed
and Roberts using a phenomenological Lagrangian
approach.\cite{Roberts97} It is important to note that the
calculations performed to date have been integrated over $4\pi$,
whereas the experimental data have been measured only over the
coverage of the CLAS. The results of these calculations are shown in
Fig.~\ref{fig:mokeev_roberts} as the dashed curves. Calculations
including the CLAS acceptance will be available soon. In general, a
very good description of the data has been achieved. Results have also
been obtained by Mokeev {\it et al.} in a phenomenological calculation
using available information on the $N$ and $\Delta$
states.\cite{Mokeev01} Parameters of this phenomenological code have
been fitted to CLAS cross-section data for real- and virtual-photon
double-charged-pion production. The results are shown in
Fig.~\ref{fig:mokeev_roberts} as the solid lines. The CLAS acceptance
was taken into account in this calculation. Neither model has yet been
adjusted to the polarization data, and therefore these results are
preliminary. There clearly is room for improvement in the model
parameters. In fact, current studies have indicated a strong
sensitivity of the helicity asymmetries to the relative contributions
of various isobaric channels and interference among
them.\cite{Strauch03,Mokeev01} These data will therefore prove to be
an important tool in baryon spectroscopy.

\end{document}